# The Energy-Entropy Principle

Rodrigo de Abreu

Centro de Electrodinâmica e Departamento de Física do IST
Lisboa, Portugal

**ABSTRACT**

In this paper, through a criticism of what we call the paradigmatic view of thermodynamics, we aim at showing a new perspective attained in this matter. The generalization of heat as internal energy (generalization of the kinetic energy concept of heat) permits the generalization of the Kelvin postulate: "It is impossible, without another effect, to convert internal energy into work" (no reference to heat or to heat reservoir).

## Introduction

The conceptual foundations of thermodynamics present subtle and interesting difficulties. The conceptual works of Caratheodory [1] and Callendar [2] are good examples of that at the beginning of the XX century. A more recent work defines the generalized view that has survived to this day the well known book of E. Fermi, Thermodynamics [3]. We call that view the paradigmatic view of thermodynamics. In this paper the method to introduce and explain the approach proposed is related to the above mentioned book in so far as its fundamental concepts are criticized and the book can therefore be regarded as a paradigm, although with some peculiarities. However we are not going to criticize especially E. Fermi's book but the paradigmatic physical approach of the middle XX century. In the text we refer to Fermi's book as (F.p.x), giving the number of the page x. This has the obvious advantage of making it easier to pass from that view to the new approach proposed, with simultaneous corroboration. The use of a well known book with physical interpretations, like Fermi's, permits to achieve this goal. In fact, Fermi sacrifices the logic consistency of the phenomenological view to the obvious interpretation emerging from the kinetic view. In fact the approach proposed in this article results from the simple interpretation based on the kinetic view used by Fermi. For an ideal gas the internal energy can be considered only kinetic. But if the ideal gas is submited to a gravitational field the internal energy has also a gravitational potential term. However the interpretation subsist for this case if we consider the internal energy and not only the kinetic energy. The generalization for other cases it is obvious and gives the hint for the changes in the phenomenological approach proposed.

In point 1 it is questioned if thermodynamics is an autonomous physical domain supported by the heat concept. This is a fundamental question because between two equilibrium points the existence or non existence of heat is dependent on the concept of heat in association with the concept of System as the ensemble of all sub-systems[4]. In fact we can separate the energy interaction terms into a work term (gravitational potential energy change or equivalent [5]-for example the energy of a piston, kinetic and potential) and an internal energy (Heat I) variation term (for example the energy of the atoms of a gas, kinetic and potential). The energy exchange between two parts of the System, two sub-systems, may be classified as heat exchange (Heat II) when one of those sub-systems is a heat reservoir (we consider a heat reservoir a large mass with constant volume in interaction with a sub-system, exchanging heat with that sub-system (Fig. 1)). This heat (exchange) is not equivalent to the classification of the internal energy term as heat (Heat I). We can have an exchange between work and internal energy and we can say that the internal energy of the System increases by a quantity equal to the work energy term - the work is transformed into heat (Heat I). The heat exchange between two sub-systems (Heat II) is not equal, of course, to the internal energy increase of each one of the sub-systems because of the existence of the work term.

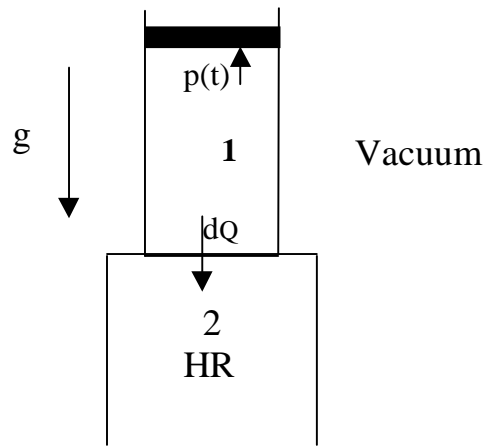

Fig. 1 The heat reservoir (HR) is sub-system 2. The other sub-system, 1, exchanges energy, exchanges heat with HR and exchanges energy, exchanges work with the piston. The internal energy of the System is the internal energy of the two sub-systems $U_1+U_2$. If we consider the atmosphere as sub-system 2 we must be aware of the adiabatic piston controversy (see 2.3.4).

If we use the word heat as internal energy (Heat I) it is possible to generalize the ordinary enunciation of the "Second" Principle of Thermodynamics and also eliminate the essential character apparently permitted by "the "First" Principle of Thermodynamics to the quantity heat, $dQ$, formally introduced in the "First" Principle.

The use of the word heat has been more recently (once again) proposed for the quantity entropy (Heat III) [2, 6]. If we use the word heat for the internal energy we have most of the properties pointed out by H. Fuchs [6] with no need to affirm that we have a heat increase in a free expansion of a gas and the word heat can obviously be used in the energetic sense emerging from the fundamental work of Joule [7, 8].

Terminological conflicts are also in the origin of the mistake pointed out in 2 and called the paradigmatic error of Thermodynamics. In 2.1 we begin by analyzing the confusion between the concept of quasi-static transformation and reversible transformation.

In 2.2 we clarify the notion of work for a reversible transformation.

In 2.3 the problems analyzed in 2.1 and 2.2 are related to the energy conservation principle and the concepts of heat - the meaning of the "First" Principle of Thermodynamics is analyzed and three examples of the conceptual difficulties emerging from the paradigmatic view are referred to at 2.3.1, 2.3.2 and 2.3.3.

In 2.3.1 the meanings of specific heats are related to the Energy Conservation Principle avoiding the paradigmatic error.

Since the concept of temperature can be derived from energy and entropy [9, 10] and related to the ideal gas [11, 12], it is possible to consider temperature a derived and not essential concept and also eliminate the tautological association between the Kelvin temperature and the ideal gas absolute temperature. This is referred in 2.3.2.

In 2.3.3 we relate the notion of adiabatic transformation to the entropy variation and to the "First" Principle.

In 2.3.4 we deal with the meanings of heat and propose the use of the word heat as internal energy. For a transformation in contact with a heat reservoir the heat exchange (Heat II) corresponds to the heat variation of the reservoir (Heat I). The essential and general character apparently given by the formal expression of the "First" Principle to the quantity $dQ$ different from the internal energy, is eliminated. In fact the use of the formal expression of the "First" Principle together with the confusion between the concept quasi-static transformation and reversible transformation originates the paradigmatic error analyzed at 2. and leads to another "subtle error" considered in another article and referred to in 2.3.

In 3. a new perspective is acquired for the "Second" Principle from the internal energy conceptualization of heat - the entropy concept can be derived from the equilibrium tendency [13] and from the asymmetric heat variation [14]. Instead of a "$1^{st}$" and a "$2^{nd}$" Principle we propose an energy-entropy principle [14].

In 4. the energy-entropy principle is interpreted microscopically in a straightforward way.

In the text the words System and sub-system are used. The System is the ensemble of all sub-systems that intervene at the transformation. To remark that we use the word system (as the ensemble of all sub-systems and not as a sub-system) written with capital letter. This being so, the transformation is always adiabatic for the System.

# 1. What is Thermodynamics?

In the introduction of his book Fermi tries to define the thermodynamics domain as the domain of heat. Curiously and paradoxically, this introduction displays the ordinary criticism to Carnot's conceptualization of heat. This generalized criticism is paradoxical because as is well known and accepted, Carnot's work is the origin of the essential concept of thermodynamics - the entropy concept (see Appendix).

As we will see, the phenomenological interpretation of heat is apparently inconsistent with other interpretations (the kinetic energy interpretation [5] or the generalization of it - the internal energy interpretation proposed).

Although Fermi considers thermodynamics (F.p.15) a heat science, in this introduction he says that thermodynamics may be considered a special branch of mechanics [13], the statistical mechanics (F.p.IX). Then if we think that thermodynamics has a non mechanical concept, like heat, as Fermi claims, the paradox is evident. This false conflict between mechanics and thermodynamics, if accepted, makes thermodynamics a mysterious physical subject. The mystery, however, can be avoided.

## 2. The Paradigmatic Error

## 2.1 The confusion between the concept of quasi-static transformation and reversible transformation

One of the basic concepts of thermodynamics is the concept of reversible transformation as a transformation with intermediate equilibrium points (F.p.4). This conceptualization is the origin of what we call the paradigmatic error [15, 16]. One example is enough to characterize the problem. In fact if we idealize a free expansion of a gas (the System) with intermediate equilibrium points [17], it is impossible to make the gas, the System, return to its initial condition. This being so, the transformation satisfies the equilibrium condition but nevertheless it is impossible for the gas to return to the initial state - the transformation is irreversible. However, if we consider another System composed of the gas and a thermal reservoir, we can apparently have a return to the initial state. After free expansion the gas is compressed and forced back to the initial conditions. The energy introduced into the gas by the compression is exchanged with the heat reservoir, as long as the thermal reservoir is big enough to hide the impossibility for the actual System (sub-system gas + sub-system reservoir) to return to the initial state. However only in the limit when the mass of the reservoir is infinite can the gas return to the initial state but even in this case the reservoir itself cannot, because the energy of the reservoir is altered. Although the gas, after free expansion, has returned to its initial state due to the compression in contact with the thermal reservoir, this reservoir retained the energy corresponding to the work done during the compression. The reservoir may obviously exchange energy with another sub-system and return to its initial state. But the new System in contact with the reservoir will, also obviously, be prevented from returning to its initial state, unless it enters into contact with another subsystem. And so on and on.

For the System (ensemble of all sub-systems) the transformation is irreversible because the System will never be able to go back to its initial state. Because of what has been said, it is essential to distinguish between the whole (System) and the parts (sub-systems). In the text the word system is used in this global sense, not in the common, currently used sense, of sub-system (the word System, with capital S, we repeat, has been used to call attention to this). Only if the exterior force is equal to the interior force and does a work [5, 18, 19] (with an energetic meaning) permitting the return to the initial conditions do we have, of course, a reversible transformation (see 2.2).

To affirm the quasi-static condition is not enough to define the reversible condition.

## 2.2 The work expression for a reversible transformation

The energy conservation principle and the tendency to equilibrium permits to write the expression (F. p.11)

$$W = \Delta U \qquad (1)$$

where *W is* the work (equivalent to the rise or fall of a weight) [5, 20] and $\Delta U$ is the System internal energy change between two equilibrium points (Fermi used for work the symbol $L$ and the sign convention is such that $L = -W$).

For a reversible transformation we can write $dW=-pdV$ where $dW$ is the infinitesimal work and $p$ and $dV$ are the equilibrium pressure and the infinitesimal volume change. In this case, and only in this case, can we write

$$dW = -pdV = dU \qquad (2)$$

Since the internal energy is also an entropy function *(U=U(V, S))* [10, 14] we have

$$dU = \left(\frac{\partial U}{\partial V}\right)_S dV + \left(\frac{\partial U}{\partial S}\right)_V dS \qquad (3)$$

*So*

$$dU = -pdV + TdS \qquad (4)$$

where $T$ is the Kelvin temperature and $p$ and $S$ are respectively the pressure and the entropy (see 3.).

For a reversible transformation the infinitesimal entropy change of the System is zero *(dS=0)*.

If we have an infinitesimal irreversible change we have (4) with *dS>0* (see 3.) ,

$$dW = dU = -pdV + TdS \qquad (5)$$

and

$$dW \neq -pdV. \qquad (6)$$

For an irreversible infinitesimal change the work is not equal to *-pdV*.

The paradigmatic error is to write $dW=-p\,dV$ for all quasi-static transformations. Only for the reversible transformations this is correct. Formally, of course, we can write $dW=-pdV$ but if we insist on calling this expression the work term we are committing what we call the paradigmatic error [l4, 15, 16, 21].

## 2.3 The "First" Principle of Thermodynamics

Fermi (F.p.4) gives to the reversible transformation the quasi-static definition (see 2.1). Only for the reversible transformation, as we saw (2.2), is the work term expressed by $dW=-pdV$.

This is a source of confusion because we have an expression and an operational means to calculate by integration *W*, a term which does not generally correspond to the energetic work.

In chapter 2 (F.p.11), Fermi says that the "First" Principle is essentially the energy conservation principle for "thermodynamic systems". If the "First" Principle is only the energy conservation principle there is no problem. But this is not so if the "First" Principle introduces a new quantity *dQ*. In fact Fermi begins by introducing the internal energy concept for "mechanical ystems", as he says (F.p.13). For "mechanical systems" he concludes that $W=\Delta U$ and he notes that the existence of energy imposes that *W* (the work term) be independent from the trajectory (F.p.12). But in two pages (F.p.l4) he says that this is not so because he affirms that thermodynamics belongs to a non-mechanical domain characterized by the heat exchange *dQ* (F.p.14 and F.p.15). (The exact sense of the word system has been previously explained in order to avoid such classifications, i.e., mechanical and thermodynamical. In this way, System corresponds to vhat Fermi designates as "mechanical system", and sub-system to what he designates as "thermodynamical system").

If we have a System with a great number of particles we can have a tendency to equilibrium if we change the exterior conditions [l3]. Then we can write between wo equilibrium points

$$W = \Delta U = \Delta U_i + \Delta U_e \qquad (7)$$

for a System composed of the sub-systems i (interior) plus a thermal reservoir (sub-system e (exterior)).

Of course we can also write

$$\Delta U_i = W + Q \qquad (8)$$

and have

$$Q = -\Delta U_e \qquad (9)$$

Curiously enough Fermi (F.p.29) also uses the word heat in the sense of internal energy and this is the terminology proposed. In fact if we aim at analyzing with generality the means of work generalizing the concepts acquired at simple "mechanical" situations, we have difficulties. For variable mass systems the force conceptualization as *f=dp/dt* and the formal use of the "First" Principle doesn't permit an universal interpretation of the quantities involved [22, 23, 24]. Also for sub-systems in interaction through a movable wall, the separation of the energy interaction and the use of the "First" Principle, have originated "subtle errors" [21, 25, 26]. We shall therefore use quantity *Q* only in the sense of energy exchanged with a heat reservoir *(Q=-ΔU_e)* [26]. If we want to continue with quantities with physical meaning we have to be careful (see 2.3.4).

### 2.3.1 The meaning of the specific heats

Fermi (F.p.20) writes for an infinitesimal transformation introducing the specific heats $C_p$, and $C_V$

$$dU - dW = dQ \qquad (10)$$

and

$$dU + pdV = dQ \qquad (11)$$

Eq. (11) is not correct for quasi-static irreversible transformations *(dW=-pdV* for reversible transformations only). We can change the state of a simple system and write for an irreversible transformation

$$dU = -pdV + TdS = dW \qquad (12)$$

If *dV=0*

$$dU = TdS = dW = C_V dT \qquad (13)$$

and therefore

$$C_V = T\left(\frac{\partial S}{\partial T}\right)_V = \left(\frac{\partial U}{\partial T}\right)_V = \frac{dW}{dT} \qquad (14)$$

We can say that $C_V$ is the energy needed to obtain a unit temperature change for constant volume.

If $p_{ext}$ (exterior pressure) is due to a gravitational field or equivalent (like the pressure due to the weight of the piston) then we have

$$dU = -pdV + TdS = dW = dW'+dW'' = C_p dT + (-p_{ext} dV) \qquad (15)$$

For $p_{ext} = p$ (*p* is the pressure at any equilibrium point),

$$C_p dT = TdS = dW' \qquad (16)$$

corresponding to the energy necessary to raise the temperature by a unit at constant pressure.

The experimental methods to determine $C_p$ and $C_V$ are, in most cases, is conceived by measuring a work term. This energy can also be measured with calorimetric methods and in this case corresponds to an exchange $dQ=-dU_e$. But here we emphasize that the measurements of $C_p$ and $C_V$ are made in physical conditions that are associated with energy measurement changes of the exterior subsystem $(dU_e)$ or a work term $dW$. For the calorimetric methods these quantities must be associated with a heat exchange under such conditions the heat exchange has a clear physical meaning [26].

### 2.3.2 The concept of temperature derived from energy and entropy and the ideal gas

If we introduce the temperature concept from energy and entropy we obtain

$$T = (\frac{\partial U}{\partial S})_V \qquad (17)$$

For an "ideal gas" Fermi introduces the energy dependence exclusively on temperature from a non existent experiment. Correctly Fermi says (F.p.22) that the temperature change is small for the Joule experiment and uses this condition to identify the Kelvin temperature $T$ with the ideal gas absolute temperature *(pV=N K A; A* is the absolute temperature). This is referred by Fermi (F.p.62), and can be avoided, but here only the problem is referred to (see the references about the ideal gas in 2.3.3).

### 2.3.3 Adiabatic transformations

On page 25 Fermi affirms that an adiabatic transformation is a reversible transformation for a thermally isolated system *(dQ=O)*. But if we are not careful we can commit the paradigmatic error. In fact, if we write

$$dU = dW + dQ \qquad (18)$$

with $dQ =0$ (adiabatic condition - without heat exchange (Heat II)), and if we also write

$$dU = -pdV + TdS \qquad (19)$$

we commit the paradigmatic error if we identify

$$dW = -pdV \qquad (20)$$

and

$$dQ = TdS. \qquad (21)$$

The adiabatic condition $dQ = 0$ only imposes $dS = 0$ for a reversible transformation (note that we are not saying that Fermi commits the error of affirming that an adiabatic transformation is equivalent to a reversible adiabatic transformation, because here, on page 25, Fermi clearly says that he uses the word adiabatic as reversible adiabatic).

Fermi (p.26) uses the equation $pV^k$ imposing the "reversible" (quasi-static) condition on the expansion of an atmospheric mass gas. He affirms the poor conductivity for the air but says nothing about the reversible condition. He commits the paradigmatic error as he confuses reversible with quasi-static.

An isentropic transformation ($dS = 0$) for a gas with state equation $p = \alpha u$ ($\alpha$ is a constant and $u$ is the energy density) satisfies the equation $pV^{\alpha+1} = constant$. For a non-relativistic monatomic gas $\alpha = 2/3$ (for a photon gas $\alpha = 1/3$). For normal ($p$, $T$) condition this gas approaches the classical Maxwell-Boltzmann behavior and satisfies $pV = BT$ with $B$ constant. We have a classical ideal gas (the photon gas exists together with the classical gas, also satisfying an equation $pV = BT$ but $B$ is not constant).

For an ideal gas ($p = \alpha u$) [11, 12], $dS = 0$ implies $pV^{\alpha+1} = constant$. For an adiabatic irreversible transformation at constant exterior pressure (equal to the interior final pressure) between two equilibrium points, the pressure and volume for the initial and final points can be fitted by an equation $pV^\delta = constant$ with $\delta \geq \alpha+1$. If the final pressure $p_2$ is not to different from the initial pressure $p_1$ ($p_2 \cong p_1$), then $\delta \cong \alpha+1$. Nevertheless the paradigmatic error is committed for the reason that the equation for the irreversible equation is $pV^\delta$ for the two points and not $pV^{\alpha+1} = constant$ for all the points ($dS = 0$). Although for the reason previously stated, i.e., the proximity of points 1 and 2, $\delta$ value is approximately $\alpha+1$.

### 2.3.4. Two meanings of Heat

On pg. 29, 31 and 56 Fermi uses the word heat in a different sense from that he carefully defines on pg. 17 without explicitly making the difference between these senses. He continues to use the word heat only. This can be a source of confusion. In fact on pg. 56 Fermi explicitly affirms the production of heat by friction ("since the heat comes from work and not from another part of the system..."). The origin of this mistake is the fluid-like conceptualization of heat. If water passes from one reservoir to another, the properties in the passage are still those associated with water and the mass of water is also conserved but if, for instance, the water runs a mill, the energy is not conserved. Even for a fluid there are properties that cannot be conserved - see Appendix). The "heat exchange" (Heat II) between two subsystems is not equivalent to the sub-systems heat variation - the energetic conceptualization of heat doesn't permit a fluid-like character for heat. In this sense we must avoid the terminology heat exchange in the sense of indestructible substance with intrinsic attributes as $H_2O$ has. However we can use the expression in the sense of' energy exchange between two subsystems in special conditions.

This being so, we propose the following nomenclature. The word heat is used as synonymous with internal energy. We can only have equality for the energy exchange between two sub-systems and the internal energy change of the sub-systems if there is no work. But as we had seen, when there exists work we can identify $dQ$ with $-dU_e$. Then, also in that situation (contact with a heat reservoir), we can think in terms of internal energy change of the exterior sub-system (the heat reservoir). This being so, we can say that we have "heat exchange" (Heat II) when we are describing an energetic exchange between sub-systems and one of these sub-systems is a "source of heat", and heat variation (Heat 1 variation) when we refer to the internal energy variation. The heat variation of the heat reservoir is equal to the heat exchange (with minus sign). The Heat 1 variation of the heat reservoir is equal in module to Heat II.

A rigorous and general description of the dynamical energy change is attained by the momentum-energy conservation principle but in special situations we can use with physical meaning the expressions $dW=dU+dU_e =dU-dQ$ and use the word heat in the sense I and II (in the sense of internal energy of the "system" and in the sense of internal energy variation of the heat reservoir). This physical interpretation led us to the energy-entropy principle considered below.

## 3. The energy-entropy principle - the generalization of Kelvin postulate

Fermi (F.p.30) gives one of the traditional enunciations of the "Second" principle of thermodynamics. The Kelvin enunciation:

"It is impossible, without another effect, to convert heat extracted from a heat reservoir into work".

Fermi (F.p.29) says that work can be transformed into heat but the reciprocal is not true. When he explains that "a body can always be heated by friction receiving a quantity of heat (Heat I variation) equalizing the work done, he is clearly using the word heat with a different meaning from that he had previously and carefully constructed (F.p.17) and (F.p.56). He uses the word heat with meaning I, in the sense of internal energy.

Although Fermi is not consistent with the previous definition of heat the physical interpretation is obvious if we introduce a clear terminology. Indeed we can generalize the Kelvin enunciation affirming that it is impossible to transform internal energy into work without other effects (no reference to heat or heat reservoir). The Kelvin enuntiation is a particular case of the new one.

As an example, consider a gas in a cylinder with a movable piston in a gravitational field. It is possible to transform the internal energy of the gas (Heat I) into work by increasing the volume. But if the volume returns to the initial value, then the internal energy is bigger or equal to the initial value (the equal value corresponds, of course, to the reversible transformation). For the same volume there is work transformed into heat not the contrary. As a consequence of this, the internal energy cannot be only a function of volume. Therefore we can introduce the entropy variable $U=U(V,S)$ and obtain eq. (3) and (4) [9, 10, 14].

It is interesting to note that if we have a hypothetical source of "movable cylinder-piston gas apparatus" (if the "evolution" has .originated such things...) then we can lift weights (we assume, of course, that the natural initial condition of the gas inside the cylinder has a pressure bigger than the atmospheric pressure) and with these weights at a higher level we can construct a hot source (if the weights are stopped at a lower level and the kinetic energy is transformed in microscopic energy, heat) which is continually renewed by the "movable cylinder-piston-gas apparatus" (this is analogous to the burn of a fossil combustible). Or with this apparatus it is possible to obtain electrical energy (for example) using directly the potential energy of the weights, without constructing a heat reservoir (as we do in an internal combustion engine). In this sense it is not true that two heat reservoirs are necessary to produce work [27, 28] because initially we have only one temperature. For that we must use the "combustible".

Therefore for a system we must conclude the following:

Beginning with only one temperature a transformation of internal energy into work is impossible without other alterations (obviously different of internal energy variation, or derived quantities). We can say that only transformations with increasing entropy are possible. We can transform heat into work for processes with an increase of entropy (for a heat reservoir it is impossible to transform the reservoir internal energy into work without other effects and this being so, beginning with only one temperature, a transformation with increasing entropy is necessary to have a global positive work). It should be noted that the previous statements cause energy to be a function of volume and of entropy [10, 14]. And since there can only be transformation of work into internal energy for the same volume, entropy variation should have only one sign (the positive sign was chosen). Therefore it is clear that a Carnot cycle (reversible) with global zero entropy variation does not contradict the previous affirmation because we must have an irreversible process to "renew" the hot reservoir. But in a more restrictive way, if the heat reservoir is not compensated and exist naturally (for instance a geothermal hot reservoir) we can say that only processes without negative entropy variation are possible.

This corresponds to what we propose to call the energy-entropy principle:

The internal energy is a function of entropy satisfying the following relation:

$$U(V, S_2) \geq U(V, S_1) \qquad (22)$$

The reversible transformation corresponds to $dS = 0$. If we choose $(\partial U/\partial S)>0$, $dS>0$. If so $S_2 > S_1$. In fact for a System we have from $U=U(V, S)$

$$dU = (\frac{\partial U}{\partial V})_S dS + (\frac{\partial U}{\partial S})_V dV = dW \qquad (23)$$

For a reversible transformation $dS=0$ and

$$dW = -pdV. \qquad (24)$$

Then, from (23) and (24)

$$p = -(\frac{\partial U}{\partial V})_S. \qquad (25)$$

Since $dU \geq 0$ when $V=constant$ from (24) we have

$$dU = TdS \geq 0. \qquad (26)$$

And if $T>0$ then $dS \geq 0$.

The entropy variation has two terms. A term associated with the volume, *(p/T) dV*, and the term *dU/T*. (The microscopic interpretation considers these two terms in phase space (see point 4.)).

## 4. The microscopic interpretation of the "second" principle

Fermi, on pg. 56, curiously enough since he constructs the entropy change associated with the expression *dS=dQ/T* (only true for a reversible transformation, but as "reversible" is confused with "quasi-static" this expression is erroneously applied also to an irreversible quasi-static transformation), says that when a body is heated by friction it receives a positive quantity of heat and, because heat results from work and not from another part of the "system" (sub-system), the increase of entropy is not compensated by the decrease of entropy of the other part. The explanation is this one if we use heat in the sense of the microscopic energy, the energy of atoms and radiation, the internal energy. But Fermi has defined heat with the other meaning, the meaning emerging from the "First" Principle. Without another explanation this is a source of confusion.

The explanation is this:

The work is transformed into internal energy (Heat I)

$$dW = dU. \qquad (27)$$

For an infinitesimal transformation

$$dW = dU = -pdV + TdS. \qquad (28)$$

If *dV=0*, from (28)

$$dW = dU = TdS. \qquad (29)$$

Then

$$dS = \frac{dU}{T} = \frac{dW}{T} > 0. \qquad (30)$$

The expression is *dS = dU/T* (not *dS = dQ/T*) and *dU* is the heat variation due to the friction work, indeed.

All of this can be easily interpreted microscopically but the paradigmatic error (the use of *dS = dQ/T* for a quasi-static irreversible transformation) with the confusion between Heat I and Heat II, originates several difficulties associated with the microscopic interpretation of the "Second" principle.

One of these difficulties is linked to the affirmation that the entropy of an "isolated system" can only increase (F.p.55). An "isolated system" is a System with constant energy (F.p.11). Here we have again a terminological problem if with the designation "isolated system" we means "thermally isolated system" (in addition to this, there is the confusion associated with the concept of thermal -Heat I and Heat II). Note that we are not saying that "isolated system" is equivalent to "thermally isolated system", of course (an "isolated system" is thermally isolated but the reciprocal cannot be true). Indeed we can have a System (e.g. a gas) changing energy with a piston in a gravitational field. The piston has no entropy, the energy of the piston is only kinetic and potential. The energy of the piston therefore is connected to the work term [5]. Therefore we separate that energy (the work term). The gas is a non-isolated System and the entropy change is positive. The gas has the energy of the microscopics entities and therefore has entropy.

For a non isolated System as well as for an isolated System the entropy change is positive or zero. Nevertheless for a sub-system the entropy change can be negative.

The entropy variation is due to the variation of the number of microstates.

Only a transformation with an increasing number of microstates is possible.

The energy contributes to this number of microstates and we can have a transformation of heat into work (with a decrease of the number of microstates associated to the momentum space) if there is an increase of the total number of microstates. This corresponds to the energy- entropy principle.

## Conclusions

In this paper we have shown that it is possible to eliminate some terminological conflicts still existing in thermodynamics.

Essentially, these conflicts are the following

1. The word heat is usually associated with the expression of the First Principle $dU = dW + dQ$ ($U$ is the internal energy and $dW$ and $dQ$ are the infinitesimal work and the infinitesimal heat). The quantity $dQ$ can be associated with the internal energy variation of a heat reservoir. When the transformation is adiabatic (the energy exchange with the heat reservoir is zero), the internal energy of the sub-system considered changes due to the work energy term. For the system (association of the sub-systems "system" + heat reservoir) the transformation is always adiabatic.

Therefore the essential quantity of thermodynamics is the internal energy and not the heat exchange. It is possible to call to the internal energy heat. But the internal energy is the energy of the microscopic entities, atoms and radiation, and because of that it is an entropy function. The increase of entropy results from the increase of internal energy for the same deformation variable (only as a limit that energy and entropy does not change - this is the reversible transformation). Then instead of a "First" and "Second" Principle, Thermodynamics is characterized by a single principle, the Energy-Entropy Principle. It is impossible for a machine to do work returning to the initial state for two non separate reasons: if the machine returns to the initial state the internal energy is the same and the entropy is also the same. But if the deformation variable is the same, the energy would have to be smaller for the work to be negative, which would correspond to a negative entropy variation. This is not possible. Only transformations with a positive entropy variation are possible.

The word heat cannot be used simultaneously for the internal energy of a sub-system ("system") and for the energy exchanged between this sub-system and the heat reservoir because this energy is not equal to the internal energy variation of the "system" but rather equal to the internal variation of the heat reservoir. This is one of the terminological conflicts that has been solved [29].

2. The second conflict, although different, is intimately connected with the first. A quasi-static transformation is not equivalent to a reversible transformation. This confusion is related to the former in the following way:

The "First" Principle

$$dU = dW + dQ \qquad (31)$$

is erroneously related to the expression

$$dU = -pdV + TdS \qquad (32)$$

derived from $U = U(V,S)$. For an irreversible quasi-static transformation these two expressions can be formally related in the following way:

$$dW = -pdV \qquad (33)$$

and

$$dQ = TdS. \qquad (34)$$

For a reversible transformation $dS=0$, and then, $dQ=0$. For an irreversible quasi-static transformation

$$dS = \frac{dU + pdV}{T}. \qquad (35)$$

If $dV=0$ then $dS = dU/T$ and not $dS = dQ/T$. It is important to note however, that, the paradigmatic error (ironically) leads to the following: people think they are applying $dS = dQ/T = dU/T + pdV/T$ to a "reversible" transformation, when they are, in fact, applying it to an irreversible, quasi-static transformation (because the quasi-static transformation is erroneously called reversible). In this irreversible quasi-static transformation

$$dS = \frac{dU + pdV}{T} \qquad (36)$$

and

$$dS \neq \frac{dQ}{T}. \qquad (37)$$

Thus is understood and eliminated the conflict between the internal energy conceptualization of heat and the heat exchanged emerging from the "First" Principle. In fact the energy conservation principle precedes the expression $dU = dW + dQ$. For a "mechanical" system (Fermi p.11)

$$dU = dW. \qquad (38)$$

But we can write for a "system" (a "thermodynamic system" in contact with a heat reservoir (this heat reservoir is a sub-system exterior to the "system")

$$dW = dU + dU_e. \qquad (39)$$

and the "mechanical" condition is satisfied.

The formal expression of the "First" Principle can induce thinking that the quantity $dQ$ is the fundamental element of Thermodynamics. This is not so and originates subtle mistakes. In this sense it is important to think without attributing to the "First" Principle wrong physical interpretations. This can be done with the construction of a single energy-entropy principle.

The microscopic interpretation of the energy-entropy principle is straightforward related to the "phenomenological" approach proposed. The tendency to a new equilibrium point with bigger entropy results from the change of volume and the change of energy, we can say with the increase of the phase space volume.

## Appendix

Carnot [30] reasons by analogy with a machine that works at the expense of water passing from a higher to a lower level reservoir.

If the water could, by itself, pass into a higher level we would have a perpetual machine. Carnot associates the heat (or caloric) passing between two sources at different temperatures with the water falling between the two levels. The quantity of water falling to the lower level reservoir is equal to the one coming out of the higher level reservoir. However, the energy of the water is not the same if the water has done work. There is, therefore, no energy conservation for the water. In the same way, there would be no energy conservation for the caloric - if the caloric existed. In fact the relevant property in this analysis is energy. So the analysis would be correct even if a substance associated with the process existed. If the water could, by itself, (at the expense of its internal energy) pass from the lower into the higher level reservoir, a perpetual machine would exist. Carnot's analysis is correct [31] and it contains the energy-entropy principle.

Today the the meaning and validity conditions of The "First" and "Second" Principle is reconsidered again [32, 33]. The physical interpretation, the meaning and validity of this interpretation, depends (as usual) on the meaning of the physical quantities considered fundamental. The simple interpretation adopted in this article results from Carnot and Boltzmann's analysis with the help of Fermi intuition. It seems convincing. It seems that the construction of a perpetuum mobile is impossible. However it is important to be aware about the claims that this is not so. Those are important and extraordinary claims. If experimentally confirmed, they are going to change the world. Another important problem linked with the "Second" Law that deserves increasing scientific interest is the problem of the origin of Life on Earth. Although it is commonly accepted that the biological entities survive (evolve) in accordance to the Energy Conservation Principle + "Second" Principle there is no consensus about the origin of Life on Earth based only on those Principles.